\documentstyle[12pt]{article}
\begin{document}
\vspace{1.cm}
\begin{center}
\    \par
\    \par
\    \par
\     \par

          \bf{  THE OBSERVATION OF COLLECTIVE EFFECTS IN CENTRAL C-Ne AND
           C-Cu COLLISIONS  AT A MOMENTUM
                OF 4.5 GeV/c  PER NUCLEON }
\end{center}
\   \par
\   \par
\   \par
\   \par
\    \par
\par
L.Chkhaidze, T.Djobava, G.Gogiberidze$^{*}$, L.Kharkhelauri\par
High Energy Physics Institute, Tbilisi State University,\par
University St 9, 380086 Tbilisi, Republic of Georgia\par
Fax: (99532) 99-06-89;  E-mail: djobava@sun20.hepitu.edu.ge or\par
\hspace{6.5cm}                     ida@sun20.hepitu.edu.ge \par
$^{*}$  Institute of Physics of the Georgian Academy of Sciencies,\par
380077 Tbilisi, Tamarashvili Str.6, Republic of Georgia
\pagebreak
\begin{center}

                     \bf{ ABSTRACT }
\end{center}
\ \par
\par
    The transverse momentum technique is used to analyse  charged-particle
exclusive data in the central C-Ne and C-Cu interactions at a momentum of
4.5 Gev/c per nucleon.
The results are presented
in terms of the mean transverse momentum per nucleon projected onto the
estimated reaction plane $<P_{x}\hspace{0.01cm}^{\prime}(Y)>$ as a function
of the
rapidity $Y$ in the laboratory system.
The observed dependence
 shows the typical $S$-shape behaviour reflecting the presence
of flow effects. The value of flow $F$ is obtained that increases
with the atomic number of the target. The Monte Carlo cascade Quark Gluon String
Model (QGSM) is used for the comparison with the experimental data. The QGSM
reproduces the spectra and the mean kinematical characteristics of the
protons ($<Y>$....$<N_{p}>$) but
underestimates their transverse flow.
\pagebreak
\par
During the last decade considerable efforts, both experimental and
theoretical, have been devoted to the study of heavy-ion collisions (HIC).
The observables investigated include produced pions, kaons, hyperons
as well as emitted nucleons and light and heavy fragments. The goal
in these experiments is to study nuclear matter under extreme conditions
of high density and temperature, i.e. to learn more about the nuclear
equation of state (EOS) [1].
\par
A signature of the compression effects predicted by the calculations
using a nontrivial EOS is the collective flow of the nuclear matter
in the expansion phase. At high energies the interaction is dominated
by two-body collisions and the collective flow  can be considered as
the consequence
of the pressure buildup in the high density zone through the short
range repulsion between nucleons, i.e. through compressional energy.
This effect leads to characteristic azimuthally asymmetric sidewards
emission of the reaction products.
The efforts to determine the EOS and the more general aspect of producing
high-energy densities over extended regions have led to a series of
experiments to study relativistic nucleus-nucleus collisions
at BEVALAC (Berkeley), GSI-SIS (Darmstadt), JINR (Dubna),
AGS (Argonne National Laboratory) and SPS (CERN). Using
the transverse momentum  analysis technique developed by $P$.Danielewicz
and G.Odyniec [2], nuclear collective flow has already been observed
for protons, light nuclei, pions and $\Lambda$ - hyperons emitted in
nucleus-nucleus
collisions at energies 0.4$\div$1.8 GeV/nucleon of BEVALAC, GSI-SIS [3-9],
and at 11$\div$14 GeV/nucleon of AGS [10,11]. The discovery
of collective sidewards flow in Au+Au at the AGS was a major
highlight at 1995 [11]. All the flow data
on asymmetric as well as symmetric nuclear collisions are reprodused
by a BUU (Boltzmann-Uehling-Uhlenbeck) transport model [12] taking a momentum
and density dependent optical potential (NMDYI - nuclear momentum
dependent Yukawa interaction) [13].
This is the first time that nuclear collective
flow in nuclear collisions can be explained quantitatively in terms of
well-known nuclear interactions.
\par
In this article we present experimental results obtained from the in-plane
transverse momentum analysis for protons in central C-Ne and C-Cu
interactions at a momentum of 4.5 GeV/c per nucleon (E=3.7 GeV/nucleon)
with the SKM-200 set-up of JINR. The signature for collective flow
had been obtained. It shows the persistence of collective flow phenomena
all the way up to AGS energies. The observed results provide a very
interesting extension of the experimental data available up to 2 GeV per
nucleon from BEVALAC and GSI-SIS on one hand and on the other
allow to bridge the
gap to the AGS energy regime and provide quantitative information on
the transverse flow and its dependence on beam
energy and projectile/target mass.
\par
 SKM-200 consists of a 2 m streamer chamber, placed in a magnetic field
of $\sim$ 0.8  T, and a triggering system. The streamer chamber was exposed to
beams of He, C, O, Ne and  Mg nuclei accelerated in the synchrophasotron
up to a momentum of 4.5 GeV/c per nucleon. The solid targets
in the form of thin discs with 0.2$\div$0.4 g/cm$^{2}$
thickness were mounted
within the fiducial volume of the chamber. Neon gas filling of the chamber
also served as a nuclear target. The central trigger was selecting events
with no charged projectile spectator fragments (with $P/Z>3$ GeV/c) within
a cone of half angle $\Theta_{ch}$ = 2$^{0}$. The
ratio  $\sigma_{cent}$/$\sigma_{inel}$  (characterizes the
centrality of selected events) is  - (9$\pm$1)$\%$ for C-Ne
and (21$\pm$3)$\%$ - for C-Cu.
Details of the experimental procedures
and data-acquisition techniques have been presented in previous publications
[14,15]. Average
measurement errors of the momentum and production angle determination
for protons are $<\Delta P/P>$= (8$\div$10)$\%$, $\Delta$$\Theta$ =1$^{0}$$\div$2$^{0}$.
\par
The data have been analysed event by event using the transverse momentum
technique of P.Danielewicz and G.Odyniec [2]. They have proposed
an exclusive way to analyse the momentum contained in directed sidewards
emission and present the data in terms of the mean transverse momentum
per nucleon in the reaction plane $<P_{x}(Y)>$ as a function of the rapidity.
After removing autocorrelation effects this method is sensitive to the true
dynamic correlations and has led to indications for collective flow effects.
 In the transverse momentum analysis the reaction plane is estimated, for each
particle $j$, with  the use of the following vector
which is constructed from the
transverse momenta $P_{{\perp}i}$ of the other particles in the same event:
\   \par
\begin{center}
$\overrightarrow{Q_{j}}$=$\sum\limits_{i\not=j}\limits^{n}$$\omega_{i}$
$\overrightarrow{P_{{\perp}i}}$ \hspace{6cm}  (1)
\end{center}
 Pions are not included. The weight $\omega_{i}$
is taken as 1 for y$_{i}$$>$ y$_{cm}$ and -1 for y$_{i}$$<$ y$_{cm}$,
where y$_{cm}$ is c.m.s. rapidity and y$_{i}$ is the rapidity of
particle $i$. The reaction plane is the plane containing
$\overrightarrow{Q_{j}}$
and the beam axis.
 The transverse momentum of each
particle  in the estimated reaction plane is calculated as
\  \par
\begin{center}
$P_{xj}\hspace{0.01cm}^{\prime}$ = $\{ \overrightarrow{{Q_{j}}}\cdot
\overrightarrow{P_{{\perp}j}}$ $/$
$\vert\overrightarrow{{{Q_{j}}}}\vert\} $ \hspace{4.1cm}  (2)
\end{center}
\par
 The average transverse momentum $<P_{x}\hspace{0.01cm}^{\prime}(Y)>$ is obtained by averaging over all
events in the corresponding intervals of rapidity.
Autocorrelations are removed
by calculating $\overrightarrow{Q}$ individually for each
particle without including that
particle into the sum (1). We defined the reaction plane for the participant
protons i.e. protons which are not fragments of the projectile ($P/Z>3$ GeV/c,
$\Theta < 4^{0}$) and
target ($P<0.2$ GeV/c). They represent the protons
participating in the collision.
The number of events and
the average multiplicity of analysed  protons $<N_{p}>$ in C-Ne and
C-Cu interactions are listed in Table 1.
\par
For the event by event analysis it is necessary to perform an
identification of $\pi^{+}$ mesons, the admixture of which
amongst the charged positive particles is about (25$\div$27) $\%$ . The
identification has been carried out on the statistical basis using
 the two-dimentional ( $P_{\parallel}$, $P_{\perp}$ ) distribution.
It had been assumed, that  $\pi^{-}$  and $\pi^{+}$ mesons
hit a given cell of the plane
( $P_{\parallel}$, $P_{\perp}$ ) with equal probability.
 The difference in multiplicity of $\pi^{+}$  and $\pi^{-}$
in each event was required to be no more than 2. After
this procedure the admixture of $\pi^{+}$ is not exceeding
(5$\div$7)$\%$.
 The temperature of the identified protons agrees with our
previous result [16], obtained by the spectra subtraction.
\par
As we study an asymmetric pair of nuclei, we chose to bypass the
difficulties associated with the center-of-mass determination and carried
out the analysis in the laboratory frame. We have replaced the
original weight $\omega_{i}$, by the continuous function
$\omega_{i}$= $y_{i}$ - $<y>$ as in [8], where  $<y>$ is the
average rapidity, calculated for each event over all the participant protons.
It is known [3], that the estimated reaction plane
differs from the true  one,
due to the finite number of particles in each event.
The component $ P_{x}$ in the true reaction plane is systematically larger
then the component $P_{x}\hspace{0.01cm}^{\prime}$
in the estimated plane, hence
\   \par
\begin{center}
$<P_{x}>$=$<P_{x}\hspace{0.01cm}^{\prime}>/<cos\varphi>$ \hspace{7.2cm}  (3)
\end{center}
were $\varphi$ is the angle between the estimated and true planes.
The  correction factor
$K$=1 $/$ $< cos\varphi >$ is subject to a large uncertainty,
especially for low multiplicity.
According to [2], for the definition of $< cos\varphi >$ we divided
randomly each event into two equal sub-events,
constructed vectors $\overrightarrow{Q_{1}}$ and
$\overrightarrow{Q_{2}}$ and estimated azimuthal angle  $\varphi_{1,2}$
between these two vectors. $< cos\varphi > $ = $< cos(\varphi_{1,2}/2) >$.
The data did not allow to perform the analysis for
different multiplicity
intervals, therefore we defined the correction factors $K$, averaged
over all the multiplicities. The values of $K$  are listed in Table 1.
 For the estimation of $< cos\varphi > $ we applied
also the different method [3], which does
not require the division of each event into two sub-classes.
\begin{center}
$< cos\varphi > $ $\approx$ $<\omega P_{x} >[<W^{2}-W>/<Q^{2}-
\sum (\omega_{i} P_{{\perp}i})^{2}>]^{1/2}$      \hspace{1cm}  (4)
\end{center}
where $W$=$\sum \vert\omega_{i} \vert$. These two methods yield consistent
results within the errors (Table 1).
\par
Fig 1,2 show the dependence of the estimated $<P_{x}\hspace{0.01cm}^{\prime}(Y)>$
on $Y$ for protons
in C-Ne and C-Cu collisions. The data exhibit the typical $S$-shape behaviour
which demonstrates the collective transverse momentum transfer between the
forward and backward hemispheres.
\par
From the mean transverse momentum distributions we can extract two main
observables sensitive to the EOS. One of them is the mean transverse
momentum averaged for positive values of rapidity
$<P_{x}>_{y>0}$.
A somehow equivalent observable is the transverse flow $F$, i.e. the slope
of the momentum distribution at midrapidity. It is a measure of the amount
of collective transverse
momentum transfer in the reaction. Technically $F$ is obtained by fitting
the central part of the dependence of
$<P_{x}\hspace{0.01cm}^{\prime}(Y)>$ on $Y$
with a sum of first and third order polynomial function.
 The coefficient of the first order term is the flow $F$.
The fit was done for Y between 0.4 $\div$ 1.9 for C-Ne and 0.2 $\div$ 2 for C-Cu.
The straight lines in Fig.1,2 show the results of this fit for the experimental
data.
The values of $F$ are listed in Table 1. The value of measured flow
$F$ is normally less than the true value because
 $P_{x}\hspace{0.01cm}^{\prime} < P_{x}$.
 The obtained values of $F$ can be considered
as lower limits of the nuclear flow in
C-Ne and C-Cu collisions. We have analysed the influence of the admixture
of ambiguously identified $\pi^{+}$ mesons  on the results.
 The error in flow $F$ includes the statistical and systematical
 errors.
\par
We have obtained also the mean transverse momentum per nucleon in the
reaction plane in the forward hemisphere of the c.m. system
$<P_{x}>_{y>0}$. The estimated and corrected (multiplied on $K$ factor)
values of $<P_{x}>_{y>0}$ are listed in Table 1. The dependence of
$<P_{x}>_{y>0}$ on beam energy and target/projectile mass is
presented in Fig.3. Results on central Ar-KCl [2,4], Ar-BaI$_{2}$ [3],
Ca-Ca [5,6], Nb-Nb [6], Ar-Pb [8] collisions from the Plastic Ball,
Diogene and BEVALAC streamer chamber groups are given together with our
results for comparison. The $<P_{x}>$ rises monotonically with E$_{beam}$,
irrespective of the projectile/target configurations. For symmetric systems
 (Ar-KCl, Ca-Ca, Nb-Nb) a linear rise with beam energy from
$<P_{x}>_{y>0}$ = 50 MeV/c at 400 MeV/nucleon to 95 MeV/c at
1800 MeV/nucleon is observed. The flow in the asymmetric system Ar-Pb,
levels off above 800 MeV/n, with data on an
intermediate mass target (BaI$_{2}$) exhibiting a somewhat higher
$<P_{x}>_{y>0}$ at 1200 MeV/n.
 In Fig.3 the BUU transport
model [12] calculations for Ar-Pb are presented. In the BUU model the dynamical
mean field and momentum dependent forces are responsible for the production of
sufficient repulsion in the nuclear collisions. The BUU calculations with
the NMDYI
reproduces well the Ar-Pb experimental results [13].
 Unfortunately the BUU calculations in the energy region of 3$\div$5 GeV/n
and light systems  C-Ne and C-Cu have not been carried out.
We extrapolated the BUU results (dashed line in Fig.3) for Ar-Pb up to
energy  4 GeV/n.  One can see, that it is desirable to perform
 precise calculations
with transport BUU models in the future for our experimental conditions.
\par
Recently the EOS/TPC collaboration reported the observation of a directed flow
behaviour for protons in Ni-Cu collisions at 2 GeV/n [9]. We estimated
the values of flow from their preliminary  dependence  of
$<P_{x}\hspace{0.01cm}^{\prime}(Y)>$
on $Y$ for both the experimental and ARC cascade model generated events, and
obtained $F \sim $120 MeV/c.
This flow is smaller than our result for C-Cu (lighter system) at higher energy
E=3.7 GeV/n.
\par
An estimate of $F$ from streamer chamber data on Ar-KCl   at 800 and
1200 MeV/n gives value of 100 MeV/c in agreement with the Plastic Ball data  for Ca-Ca
at the same energy.  For the heavier system Nb-Nb [6] the flow
$F$ increases from
135 MeV/c at 400 GeV/n  to 160 MeV/c at 1050 MeV/n.
\par
One can see from the Table 1, that with the increase of the atomic number of
the target A$_{T}$, the values of $F$ and $<P_{x}>_{y>0}$ rise. A similar
tendency had been observed at lower energies [3-6]. The results from AGS [10,11]
are not presented in Fig.3, since they present  collective sidewards
flow in somewhat different  observables.
\par
The systematic study of the dependence of the flow on the target/projectile mass
and the beam energy represents a comprehensive body of data that should
enable theoretical model calculations to obtain further information on the
nuclear matter EOS.
\par
Several theoretical models of nucleus-nucleus collisions at
high energy have been
proposed [17]. In this paper the Quark Gluon String Model
(QGSM) [18] is used for a comparison with experimental data. The QGSM is based
on the Regge and string phenomenology of particle production in inelastic
binary hadron collisions [19].
The QGSM simplifies the nuclear effects (neglects the potential
interactions between hadrons, coalescence of nucleons and etc.).
A detailed description and
comparison of the QGSM with experimental data over a wide energy range can be
found in paper [20].
The procedure of generation consists of 3 steps: the definition of
configuration of colliding nuclei, production of quark-gluon strings
and fragmentation of strings (breakup) into observed hadrons.
After
hadronization the newly formed secondary hadrons are allowed to rescatter.
 In the
QGSM the sidewards flow is a sole result of the rescattering
of secondaries, which
produces the amount of collective energy.
 The model
yields a generally good overall fit to most experimental data [20,21].
\par
We have generated C-Ne and C-Cu interactions using the Monte-Carlo generator
COLLI, based on the QGSM and then traced through the detector
and trigger filter.
\par
In the generator COLLI there are two possibilities to generate events:
1) at not fixed impact parameter $b$ and 2)at fixed $b$.
 From the $b$ distributions
 we obtained the mean values
$<b>$=2.20 fm for C-Ne collisions and $<b>$=2.75 fm for C-Cu
and total samples of events for these $<b>$ had been generated.
 The QGSM overestimates the production of low
momentum protons  with $P<$0.2 GeV/c, which  are mainly the target fragments
and were excluded from the analysis.  From the analysis of generated events
the protons with deep angles greater  60$^{0}$ had been excluded, because
 such vertical tracks  are registered with less efficiency on the experiment.
 The proton multiplicity, rapidity, $<P_{T}>$ and momentum
distributions of
generated events for
fixed and not fixed $b$ are consistent and reproduce the data.
 The corresponding
values of $<Y>$ and $<N_{p}>$ are listed in Table 1. For generated events
the component in the true reaction plane $ P_{x}$ had been calculated.
The dependences of  $<P_{x}(Y)>$ on $Y$
 are shown in
Fig.1,2. For the visual presentation,
we approximated these dependences by polynoms (the curves in Fig.1,2).
From the comparison of the dependences  of $<P_{x}(Y)>$ on $Y$ obtained
by the model in  two regimes - for fixed and  not fixed $b$, one can conclude,
that the results are consistent and it seems, that in our experiment the values of
b=2.20 fm for C-Ne, b=2.75 fm for C-Cu are probable.
The QGSM yields a significant flow signature, which follows trends similar
to the experimental data, but is smaller.
 To be convinced, that the
significant sidewards deflection in Fig.1,2 (for both experiment and QGSM) is
due to correlations within the events, and can not be the result of detector
biases or finite-multiplicity effects, we obtained
the $<P_{x}(Y)>$ on $Y$ for events composed by randomly selecting
tracks from different QGSM events (within the same multiplicity range)
(Fig.1,2). One can see from Fig.1,2, that in these events there is no
correlation with reaction plane.
 The values of $F$, obtained
from the QGSM are listed in Table 1. One can see, that the QGSM underestimates
the flow at our energies, as the values of $F$ from
the experimental data are lower limits (not corrected for $K$).
 This model underestimates also the
transverse flow at BEVALAC energies (1.8 GeV/n) [2,3]. As shown by H.Stocker
and  Greiner [22], the reason that the  QGSM fails to reproduce the flow
data in the
energy region of 1$\div$5 GeV/n, is the neglect of mean-field effects. At higher
energies the influence of a mean field is expected to be weaker, so it is
believed  that the QGSM will still underestimate the flow,
but will give a better
estimate than at lower energies. The QGSM with proper
treatment of rescattering
predicts observable collective flow in Pb-Pb collisions at AGS energy
E=10 GeV/n , of the order of 125 MeV/c [18].
\par
In summary, in this paper we have reported experimental results obtained
from the analysis of central C-Ne and C-Cu collisions at a momentum of
4.5 GeV/c per nucleon with the SKM-200 setup. The data have been analysed
event by event using transverse momentum technique.
The results are presented
in terms of the mean transverse momentum per nucleon projected onto the
estimated reaction plane $<P_{x}\hspace{0.01cm}^{\prime}(Y)>$
as a function of $Y$ in the laboratory system.
The observed dependence of the
$<P_{x}\hspace{0.01cm}^{\prime}(Y)>$ on $Y$ shows the typical
$S$-shape behavior
reflecting the presence of flow effects. From this dependence
we have extracted the flow  $F$, defined as the slope
at midrapidity, $F$=109$\pm$10 (MeV/c) for C-Ne,
$F$=161$\pm$15 (MeV/c) ---  C-Cu.
 The obtained values
of $F$ can be considered as lower limits of flow.
The correction factors
$K$, due to the uncertainties on the determination of the raction plane,
 had been estimated
by two methods, which yield consistent results.
The $F$ increases
with the atomic number of target $A_{T}$, which indicates on the rise of
collective flow effect.
The values of
$<P_{x}\hspace{0.01cm}^{\prime}>_{y>0}$ and  $<P_{x}>_{y>0}$
had been obatined.\\
$<P_{x}\hspace{0.01cm}^{\prime}>_{y>0}$=74$\pm$8  (MeV/c) - C-Ne, \hspace{0.3cm}
$<P_{x}\hspace{0.01cm}^{\prime}>_{y>0}$=114$\pm$12 (MeV/c) - C-Cu.\\
$<P_{x}>_{y>0}$=97$\pm$11 (MeV/c) - C-Ne, \hspace{0.1cm}
$<P_{x}>_{y>0}$=145$\pm$18 (MeV/c) - C-Cu.\\
 The values of $<P_{x}>_{y>0}$ had been compared
with the results at lower energies of 0.4$\div$1.8 GeV/n for various
projectile/target configurations. The $<P_{x}>_{y>0.}$ increases
with the beam energy.
The Monte-Carlo Cascade Quark Gluon String Model (QGSM) was
used for the comparison with the experimental results. The QGSM
reproduces the spectra and the mean kinematical characteristics of the
protons ($<Y>$...$<N_{p}>$) but
underestimates their transverse flow.
\   \par
\   \par
ACKNOWLEDGEMENTS
\par
\   \par
We would like to thank Prof. N.Amaglobeli for  his
support. We are indebted to M.Anikina, S.Khorozov, J.Lukstins,
  L.Okhrimenko, G.Taran
for help in obtaining the data.
We are very grateful to N.Amelin for providing us with the QGSM
code program COLLI, also to
Z.Menteshashvili for many valuable discussions.
\par
The research described in this publication was made possible in part
by Grant  MXP000 from the International Science Foundation and  Joint Grant
MXP200 from the International Science Foundation and Government of Republic
of Georgia.
\pagebreak
\par

\newpage
\begin{center}
\bf{FIGURE CAPTIONS}
\end{center}
\  \par
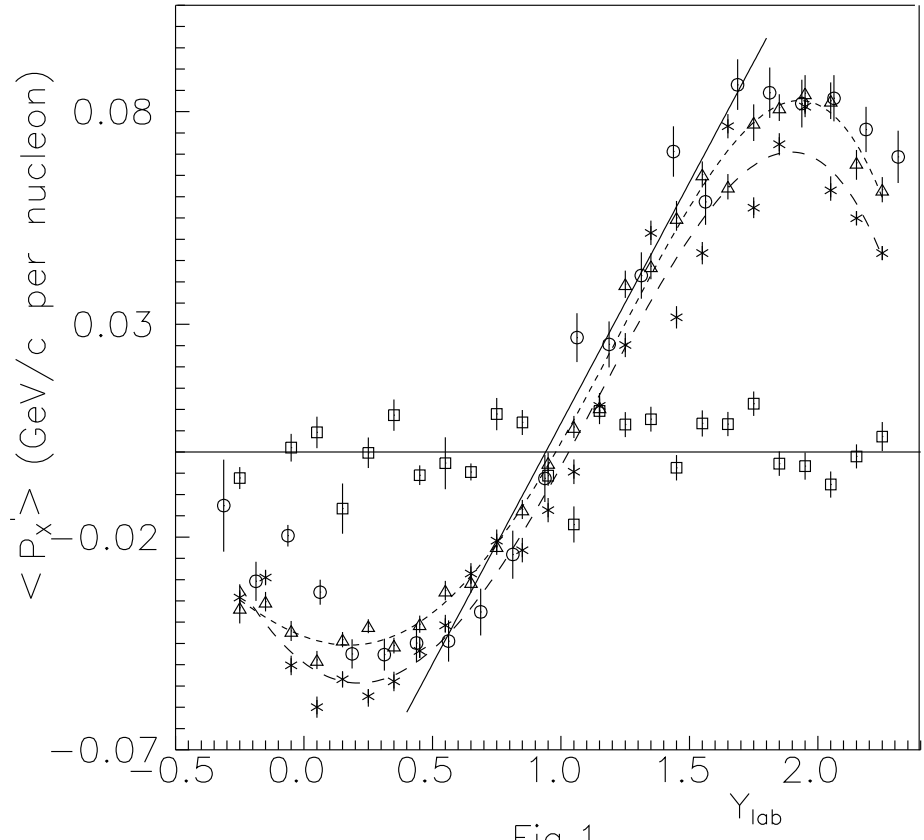
{\bf{Fig.1}}
The dependence of $< P_{x}\hspace{0.01cm}^{\prime}(Y) >$ on Y$_{Lab}$ for
protons in C-Ne collisions. $\circ$ -- the experimental data,
$\bigtriangleup$ -- QGSM generated data for fixed $b$= 2.20 fm,
$\ast$ -- QGSM generated data for not fixed $b$,
\mbox{\put(3.,0.){\framebox(6.,6.)[cc]{}}}~~~~
--  events composed by randomly selected
tracks from the different QGSM events (within the same multiplicity range).
The solid
line is the result of the approximation of experimental data by sum of
first and third order polynomial function in the interval of
 Y - 0.4 $\div$ 1.9
. The dashed curves for
visual presentation of the QGSM events (short dashes - for fixed $b$,
long dashes -for not fixed $b$)
- result of approxomation by 4-th order polynomial function.
\   \par
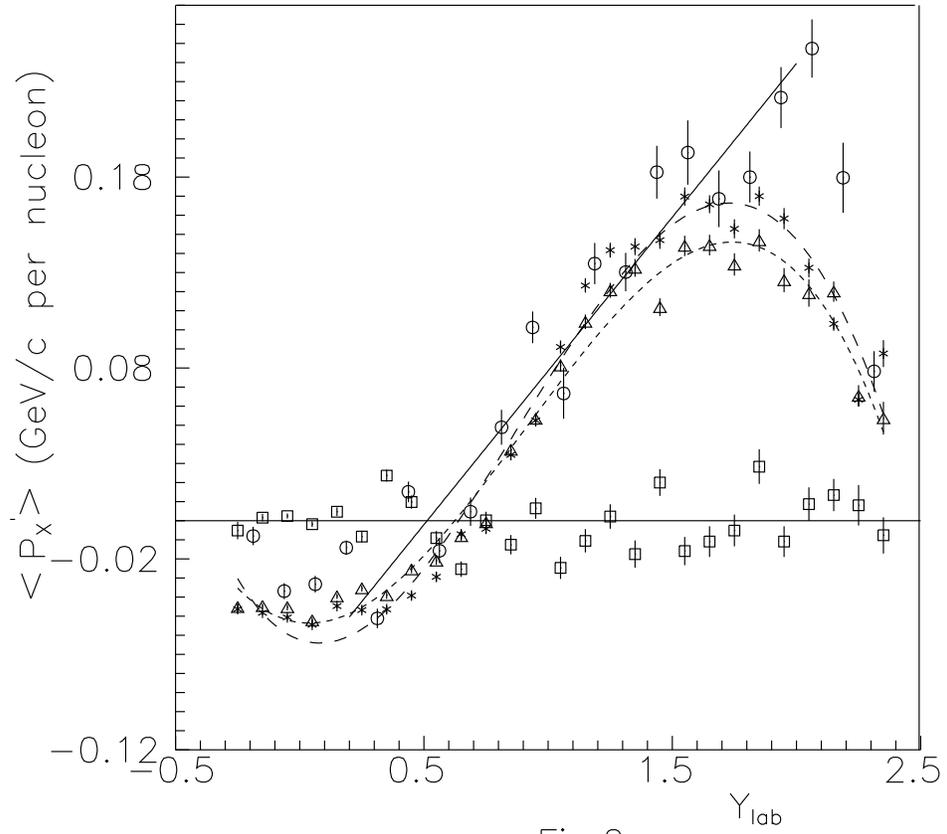
{\bf{Fig.2}}
The dependence of $< P_{x}\hspace{0.01cm}^{\prime}(Y) >$ on Y$_{Lab}$ for
protons in C-Cu collisions. $\circ$ -- the experimental data,
$\bigtriangleup$ --  QGSM generated data for  fixed $b$= 2.75 fm,
$\ast$ --  QGSM generated data for not fixed $b$,
\mbox{\put(3.,0.){\framebox(6.,6.)[cc]{}}}~~~~
--  events composed by randomly selected
tracks from the different QGSM events (within the same multiplicity range).
The solid
line is the result of the approximation of experimental data by sum of
first and third order polynomial function in the interval of
 Y -  0.2 $\div$ 2
. The dashed curves for
visual presentation of the QGSM events (short dashes - for fixed $b$,
long dashes -for not fixed $b$)
- result of approxomation by 4-th order polynomial function.
\   \par
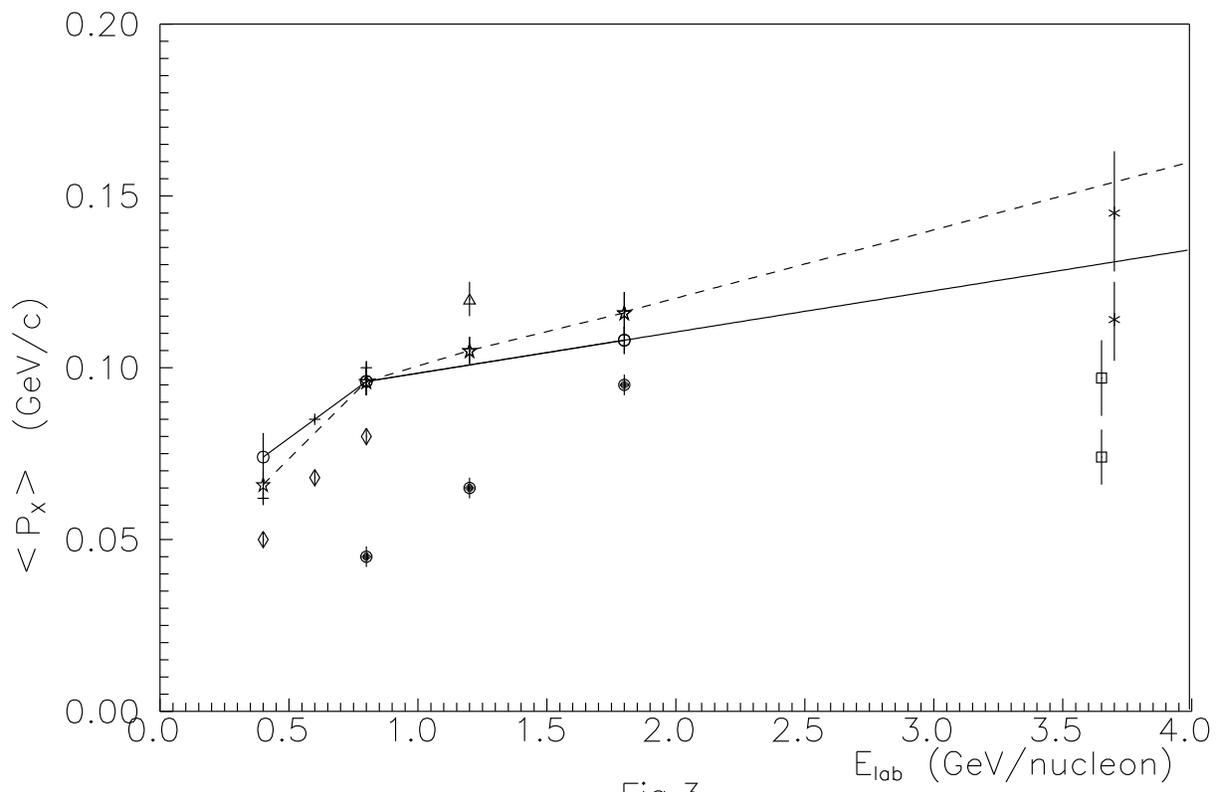
{\bf{Fig.3}}
The average transverse momentum per nucleon in the reaction plane
in the forward hemisphere of the c.m. system  as a function of
beam energy for various projectile/target configurations.
$\circ$ --  Ar-Pb [8] ,  $\bigtriangleup$ -- Ar-BaI$_{2}$ [3],
$\diamond$ -- Ca-Ca [5,6], $\bullet$ -- Ar-KCl [2,4], $\dag$ -- Nb-Nb [6],
$\star$ -- The BUU calculations for Ar-Pb [13],
\mbox{\put(3.,0.){\framebox(6.,6.)[cc]{}}}~~~~
-- C-Ne , $\ast$ -- C-Cu the estimated and multiplied on correction
factor $K$. The solid and dashed lines connect experimental and
BUU values of Ar-Pb and are extrapolated up to E=4 GeV/n.
\   \par
\   \par
\    \par
\begin{center}
\bf{TABLE CAPTIONS}
\end{center}
\   \par
Table 1. The number of experimental and QGSM simulated
events,  the average  multiplicity of participant
protons $<N_{p}>$, the mean rapidity $<Y>$,
 the correction factor $K$ ,
 the flow  $F$ and
the average transverse momentum per nucleon in the
reaction plane  in the forward hemisphere of the c.m. system
$<P_{x}>_{y>0}$.
\newpage
Table 1. The number of experimental and QGSM simulated
events,  the average  multiplicity of participant
protons $<N_{p}>$, the mean rapidity $<Y>$,
 the correction factor $K$ ,
 the flow  $F$ and
the average transverse momentum per nucleon in the
reaction plane  in the forward hemisphere of the c.m. system
$<P_{x}>_{y>0}$.
\   \par
\   \par
\   \par
\   \par
\   \par
\   \par
\begin{tabular}{|l|c|c|}    \hline
&     &   \\
\hspace{0.5cm} &\hspace{1.cm} C-Ne \hspace{1.3cm} &\hspace{1.3cm} C-Cu\hspace{1.6cm}        \\
&     &   \\
\hline
  Number of exper. events  &      723      &      305      \\
\hline
  Number of generated events, &      &                \\
  b not fixed     &   2925    &   3194    \\
\hline
  Number of generated events, &  b=2.20 fm   &  b=2.75 fm   \\
  b fixed         &   2128    &   3210     \\
\hline
    $<N_{p}>_{exp}$          &  12.4 $\pm$ 0.5   &  19.5$\pm$ 0.6   \\
\hline
    $<N_{p}>_{mod}$ not fixed b &  11.5 $\pm$ 0.2  & 21.7$\pm$ 0.3    \\
\hline
    $<N_{p}>_{mod}$  fixed b & 12.0 $\pm$ 0.2   &  22.6$\pm$ 0.3    \\
\hline
    $<Y>_{exp}$              & 1.07 $\pm$ 0.07   &  0.71$\pm$ 0.08   \\
\hline
    $<Y>_{mod}$ not fixed b & 1.03 $\pm$ 0.03   &  0.64$\pm$ 0.05    \\
\hline
    $<Y>_{mod}$ fixed b & 1.05 $\pm$ 0.03   &  0.62$\pm$ 0.05    \\
\hline
 $K$=1/$< cos\varphi>$ by method [2] &  1.27 $\pm$ 0.08 &  1.31 $\pm$ 0.04   \\
\hline
 $K$ by method [3]        &  1.42 $\pm$ 0.06 &  1.39 $\pm$ 0.04   \\
\hline
  $F_{exp}$ (lab.) (MeV/c)   &  109  $\pm$10  & 161$\pm$15     \\
\hline
  $F_{mod}$ (MeV/c) not fixed b         &   92 $\pm$8     & 164$\pm$14     \\
\hline
  $F_{mod}$ (MeV/c) fixed b   &   95 $\pm$9       & 153$\pm$13     \\
\hline
  $ <P_{x}\hspace{0.01cm}^{\prime}>_{y>0}$ (MeV/c) & 74 $\pm$8 & 114$\pm$12 \\
\hline
  $ <P_{x}>_{y>0}$  (MeV/c)    &     &      \\
  multiplied on $K$ by method [2]    &   97 $\pm$11     & 145$\pm$18     \\
\hline
\end{tabular}
\end{document}